# LfEdNet: A Task-based Day-ahead Load Forecasting Model for Stochastic Economic Dispatch

Jiayu Han, Lei Yan, *Student Member, IEEE*, Zuyi Li, *Senior Member, IEEE*

*Abstract*—Load forecasting is one of the most important and studied topics in modern power systems. Most of the existing researches on day-ahead load forecasting try to build a good model to improve the forecasting accuracy. The forecasted load is then used as the input to generation scheduling with the ultimate goal of minimizing the cost of generation schedules. However, existing day-ahead load forecasting models do not consider this ultimate goal at the training/forecasting stage. This paper proposes a task-based day-ahead load forecasting model labeled as LfEdNet that combines two individual layers in one model, including a load forecasting layer based on deep neural network (Lf layer) and a day-ahead stochastic economic dispatch (SED) layer (Ed layer). The training of LfEdNet aims to minimize the cost of the day-ahead SED in the Ed layer by updating the parameters of the Lf layer. Sequential quadratic programming (SQP) is used to solve the day-ahead SED in the Ed layer. The test results demonstrate that the forecasted results produced by LfEdNet can lead to lower cost of day-ahead SED while maintaining a relatively high forecasting accuracy.

*Indices*
| | |
|---|---|
| $g$ | Index for thermal units |
| $t$ | Index for hourly time slots |
| $l$ | Index for transmission lines |
| $m$ | Index for buses |

*Parameters*
| | |
|---|---|
| $N_g$ | Number of thermal units |
| $T$ | Number of hourly time slots |
| $P_{g,max}$ | Maximum power output of thermal unit $g$ |
| $P_{g,min}$ | Minimum power output of thermal unit $g$ |
| $RU_g$ | Maximum upward ramp of thermal unit $g$ |
| $RD_g$ | Maximum downward ramp of thermal unit $g$ |
| $k_r$ | Load factor |
| $F_l$ | Maximum power flow of line $l$ |
| $\lambda_s$ | Shortage penalty coefficient |
| $\lambda_e$ | Excess penalty coefficient |
| $\lambda_l$ | Penalty coefficient for power flow |
| $a_g, b_g, c_g$ | Coefficient of cost function of generation scheduling |

*Variables*
| | |
|---|---|
| $L_p(\cdot)$ | Loss function of the Lf layer |
| $\theta$ | Parameters of the Lf layer |
| $x$ | Input data of the Lf layer |
| $y$ | Predicted load of the Lf layer |
| $\sigma^2$ | Empirical variations of predicted load |
| $P_{g,t}$ | Power output of thermal unit $g$ at time $t$ |
| $p(y\|x;\theta)$ | Conditional probability of $y$ given $x$ |
| $E_{y\sim p(y\|x;\theta)}$ | Expectation of $y$ |
| $\Gamma_{l,m}$ | Shift factor for line $l$ with respect to bus $m$ |
| $P^*(x,\theta)$ | Optimal economic dispatch |
| $L_{t_{new}}(\cdot)$ | Task loss function |
| $I(\cdot)$ | Indicator whether a constraint is satisfied or not |
| $C^*_{p,t}$ | Optimal cost of generating schedule |
| $C_1$ | Penalty term of generation shortage |
| $C_2$ | Penalty term of power flow |
| $C_3$ | Regularization term |
| $PDF(\cdot)$ | Probability density function |
| $CDF(\cdot)$ | Cumulative distribution function |

## I. INTRODUCTION

LOAD forecasting is an important task of the operation of modern power systems. There are, roughly speaking, four types of load forecasting based on forecast horizon, including very-short-term, short-term, medium-term, and long-term load forecasting that forecasts load minutes to one-hour ahead, one-hour to one-week ahead, one-week to one-year ahead, and one-year ahead or longer, respectively. This paper focuses on the day-ahead hourly load forecasting, which is one type of short-term load forecasting (STLF) that is mainly used for scheduling generation, transmission, and distribution, and for other short-term decision making in power systems, like generation reserve, system security, financial planning, demand-side management, and so on [1]. Load forecasting methods are roughly grouped in three major groups based on when they were firstly used: traditional forecasting methods, modified traditional methods, and modern forecasting methods [2].

The most widely used traditional forecasting methods include the regression method [3], exponential smoothing [4], multiple linear regression [5], and iterative reweighted least-squares methods [6]. Modified traditional forecasting methods include auto-regressive moving average (ARMA) [7], auto-regressive integrated moving average (ARIMA) [8], the combination of several algorithms [9], and so forth. Sudden change at irregular time intervals and periods of high and low volatilities [10] create challenges for these methods.

Modern forecasting methods including various machine learning models in general and deep learning in particular have been applied to improve the accuracy of load forecasting. Support vector machine (SVM), one of the supervised machine learning models, has been used for load forecasting [11]. Techniques like particle swarm global optimization and grid traverse algorithm are combined with SVM to improve the performance of SVM for load forecasting [12-13]. Other machine learning methods used for load forecasting [14] include fuzzy regression trees [15], Kalman filtering [16], forest regression [17], etc. With the advent of the graphics processing unit (GPU), the speed of computing is highly improved and deep learning methods become more and more popular. Deep learning is part of a family of machine learning methods based on artificial neural networks with representation learning [18]. Long short-term memory (LSTM) recurrent neural network-based framework [23], deep residual networks [24], and a two-step short-term load forecasting model with Q-learning [25] are used for STLF. A recently developed technique called Bayesian deep learning is employed to make probabilistic load forecasting [26]. Convolutional neural network (CNN) technique is also introduced for load forecasting in recent years. A CNN based bagging learning approach [27], a method combined CNN and LSTM [28], and a multi-scale CNN with time cognition [29] are applied to solve the STLF problems.

Accuracy is a very important metric to evaluate a load forecasting model so existing researches mostly focus on how to improve the accuracy of load forecasting. Their main ideas are to design a better predictive model to minimize the prediction loss function. Then, the results are treated as "correct results" to be used in later optimization problems (e.g., unit commitment and economic dispatch). The advantage of this sequential approach is that the predictive model is independent of any future task so the predicted results will not be influenced by a future task. However, the loss function is different from the ultimate criterion on which the predictive models are evaluated. To be specific, the loss function of most of the load forecasting methods is to minimize the errors between the predicted load and the real load. However, for power system operations, the ultimate task is to minimize the cost of a scheduling procedure using the predicted load in most cases. The above-mentioned predictive models generally do not consider the ultimate task. In this aspect, the load forecasting results based on traditional machine learning methods or modern deep learning methods mentioned above fall short of meeting the need of practical applications. Furthermore, every predictive method does have error and if the predicted values of the load have a big bias compared to the real values, it may cause model bias and cannot provide credible generation schedule for power system operations.

To resolve the above issues, a new task-based load forecasting model LfEdNet is proposed as the main contribution of this paper. This LfEdNet has two high-level layers. The first layer ("Lf layer") is a four-layer artificial neural network (ANN) model that outputs the forecasted load to the second layer ("Ed layer"), which is a stochastic economic dispatch (SED) model that obtains the optimal dispatch results. The LfEdNet minimizes the overall loss function, which is also the loss function of the Ed layer, by updating the parameters of the Lf layer and optimizing the SED at the Ed layer. The biggest difference between LfEdNet and other forecasting models is that LfEdNet is a task-based model and its objective is to minimize the cost of day-ahead stochastic economic dispatch (simply SED cost thereafter). To achieve that objective, it combines the load forecasting model and SED model together in one deep neural network. Thus, the proposed model is a closed-loop forecasting model that considers the entire process of power system operations. Compared to other model-free load forecasting model, the results of the proposed model can guarantee not only to obtain relatively high forecasting accuracy but also to find relatively low SED cost, which is of practical significance.

The rest of this paper is organized as follows. Section II describes the two high-level layers of the proposed LfEdNet in detail. Section III presents several case studies to examine the advantages of the proposed LfEdNet. Section IV concludes this paper.

## II. Proposed Task-based LfEdNet

This section presents the architecture of the proposed task-based LfEdNet, the four-layer ANN model in the Lf layer, the mathematical formulation and the solution method for the SED in the Ed layer, and the training process of the LfEdNet.

### A. The Architecture of LfEdNet

In this paper, a task-based load forecasting model is proposed, which is called LfEdNet. It combines load forecasting and economic dispatch together as one deep neural network model. Different from other existing load forecasting models, LfEdNet is mainly focused on the tradeoff between the accuracy of load forecasting and the optimal cost of economic dispatch.

The architecture of the proposed LfEdNet is shown in Fig. 1. It contains two high-level layers including Lf layer which performs load forecasting and Ed layer which performs SED, and two passes including the forward pass and the backward pass. In the forward pass of LfEdNet, the Lf layer, which is a four-layer ANN model, forecasts the load and associated variance. Then, the Ed layer, which is an SED model, calculates the optimal economic dispatch based on the forecasted load and variance. Next, the task loss is obtained using the economic dispatch from the Ed layer. In the backward pass of LfEdNet, the derivative of the loss function with respect to the parameters in the Lf layer is obtained. Then, the parameters of the Lf layer are updated according to gradient descent methods. The iterative process between the forward pass and the backward pass will continue until certain criteria are satisfied. During the forecasting stage, given the input data, a load forecast and corresponding optimal dispatch can be obtained.

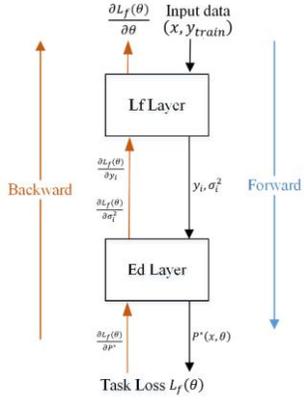

Fig. 1. The architecture of LfEdNet.

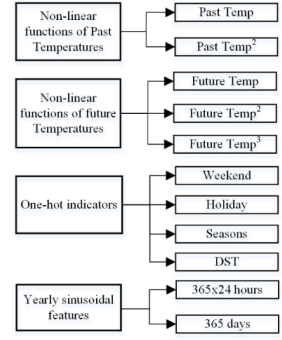

Fig. 3. Detailed features of input data in Lf layer

### B. Lf Layer

Fig. 2 shows the detailed architecture of the four-layer ANN model in the Lf layer. In addition to the main body of the four-layer ANN, the residual connection is added from the input layer to the output layer, which can avoid the problem of vanishing gradients [31]. There are three hidden layers with 250 nodes in each hidden layer. Batch normalization is added in each hidden layer, which acts as a regularization term to reduce overfitting, solves covariate shift problem, and speedups training [32]. The features of input data include historical load data, temperature, one-hot indicators, and yearly sinusoidal features, as shown in Fig. 3. For temperature, both past and predicted data are considered. Since the relationship between temperatures and load is not simply linear, in this paper, quadratic polynomial function of past temperature and cubic polynomial function of predicted temperature are used. There are four kinds of one-hot indicators used as input features including seasons, holidays, weekends, and daylight-saving times. Since sine and cosine functions are able to capture any complex form of seasonality or cyclicality, they are used in this paper to find the yearly sinusoidal features by two different frequencies, e.g. one cycle is $356 \times 24$ hours and the other is 365 days. The output of the ANN model is the forecasted electrical loads over all 24 hours of the next day. The loss function in Eq. (1) is used to pre-train and initialize the ANN model.

$$L_p(\theta) = \frac{1}{N}\sum_{i=1}^{N}(y_{train,i} - y_i)^2 \qquad (1)$$

where $N$ is the total number of the training data, $y_{train,i}$ is the actual load and $y_i$ is the forecasted load, and $\theta$ is the weights or parameters of the ANN.

The outputs of the Lf layer include the forecasted load $y_i$ and the empirical variance $\sigma_i^2$.

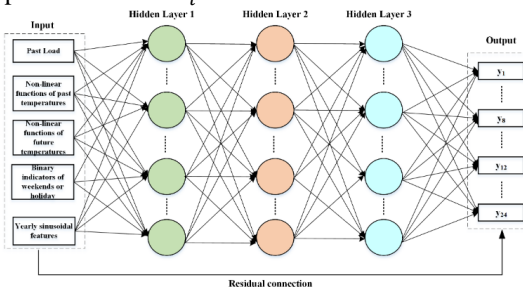

Fig. 2. Four-layer artificial neural network in Lf layer.

### C. Ed Layer

The objective of the Ed layer is to find the optimal SED cost using the forecasted load from the Lf layer. This model is inspired by Ref. [33], which proposes a task-based end-to-end model in stochastic optimization. The framework proposed in Ref. [33] is applied to this paper. The difference between this paper and Ref. [33] is that a more complex stochastic optimization model is used in this paper. The model in this paper considers a realistic SED model that includes more than one generator, generation ramping and capacity constraints, generation/load balance constraints, and transmission line constraints. In comparison, the model in Ref. [33] includes only one generator and its ramping constraint.

The SED model is built in this layer as a 'certainty-equivalent' problem, which is given by (2):

$$\min_{P \in R^{T \times N_g}} \sum_{t=1}^{T} E_{y \sim p(y|x;\theta)}(C_{p,t}) \qquad (2)$$

s.t.
$$P_{g,t} - P_{g,t-1} \leq RU_g, \forall g, t \qquad (2.1)$$
$$P_{g,t-1} - P_{g,t} \leq RD_g, \forall g, t \qquad (2.2)$$
$$P_{g,min} \leq P_{g,t} \leq P_{g,max}, \forall g, t \qquad (2.3)$$
$$E_{y \sim p(y|x;\theta)}\left(\sum_{g}^{N_g} P_{g,t} - y_t\right) = 0, \forall j \qquad (2.4)$$
$$E_{y \sim p(y|x;\theta)}\left(\left|\sum_m \Gamma_{l,m}\left(\sum_{g \in \mathcal{G}(m)} P_{g,t} - k_r \times y_t\right)\right|\right) \leq F_l \qquad (2.5)$$

where, $C_{p,j} = \sum_{g=1}^{N_g} a_g P_{g,t}^2 + b_g P_{g,t} + c_g$. $p(y|x;\theta)$ represents the probability distribution function of the forecasted load $y$ as a function of training data $x$ and the weights of the ANN, $\theta$. The objective function (2) is to minimize the total cost of all generators. Eqs. (2.1)-(2.2) are ramping up and down constraints, respectively. Eq. (2.3) limits the minimum and maximum power outputs. Eq. (2.4) is the probabilistic generation/load balance constraints and Eq. (2.5) represents probabilistic transmission line constraints. Note that $y_j$ is the forecasted load and $\theta$ is the weights of the Lf layer.

The optimal dispatch of SED model (2) is represented as $P^*(x, \theta)$. The task loss function of LfEdNet is represented as $L_f(\theta)$ shown in Eq. (3).

$$L_f(\theta) = \sum_{t=1}^{T}\left[ I\left(E_{y \sim p(y|x;\theta)}\left(\sum_{g=1}^{N_g} P_{g,t}^* - y_{train,t}\right) = 0\right) + I\left(E_{y \sim p(y|x;\theta)}\left(\left|\sum_m \Gamma_{l,m}\left(\sum_{g \in \mathcal{G}(m)} P_{g,t}^* - k_r \times y_{train,t}\right)\right|\right) \leq F_l\right) + E_{y \sim p(y|x;\theta)}(C_{p,t}^*) \right] \qquad (3)$$

where $C_{p,t}^* = \sum_{g=1}^{N_g} a_g {P_{g,t}^*}^2 + b_g P_{g,t}^* + c_g$.

In (3), task loss $L_f(\theta)$ is a function of $\theta$. $I(\cdot)$ represents the indicator whose value is zero if the constraints (2.4) and (2.5) are satisfied and infinite otherwise.

The derivative of task loss function (3) with respect to the parameter $\theta$ is complicated because of the indicator $I$. First, it should be checked whether all probabilistic constraints are satisfied or not. If yes, $\theta$ is updated by $\nabla_\theta C_{p,t}^*$. Otherwise, $\theta$ is updated by taking gradient descent of the violated probabilistic constraints. Given a large number of training data, the above gradient decent procedure is time-consuming. An effective way that can help speed up the calculation is to add violation/shortfall penalty to the objective function [34], which is a good way to remove indicator of the loss function. According to the method in Ref. [34], the modified SED model is shown as follows.

$$\min_{P \in R^{T \times N_g}} \sum_{t=1}^{T} \mathrm{E}_{y \sim p(y|x;\theta)}[C_{1,t} + C_{2,t} + C_{3,t} + C_P] \quad (4)$$

$$\text{s.t. } (2.1) - (2.3) \quad (4.1)$$

Where

$$C_{1,t} = \lambda_s \max\left(y_t - \sum_{g=1}^{N_g} P_{g,t}, 0\right) + \lambda_e \max\left(\sum_{g=1}^{N_g} P_{g,t} - y_t, 0\right) \quad (4.2)$$

$$C_{2,t} = \lambda_l \max(\sum_m \Gamma_{l,m}(\sum_{g \in \mathcal{G}(m)} P_{g,t} - k_r \times y_t) - F_l, 0) + \lambda_l \max(\sum_m \Gamma_{l,m}(k_r \times y_t - \sum_{g \in \mathcal{G}(m)} P_{g,t}) - F_l, 0) \quad (4.3)$$

$$C_{3,t} = 1/2 \times \left(\sum_{g=1}^{N_g} P_{g,t} - y_t\right)^2 \quad (4.4)$$

Different from objective function (2), the new objective function in Eq. (4) adds three more penalty terms (i.e., $C_{1,t}$, $C_{2,t}$, and $C_{3,t}$). $C_{1,t}$ is the penalty term for generation shortage and generation excess, where $\lambda_s$ and $\lambda_e$ are shortage penalty coefficient and excess penalty coefficient, respectively. $C_{2,t}$ is the penalty term of power flow and $\lambda_l$ is the corresponding penalty coefficient. $C_{3,t}$ is the regularization term which can make the generation schedules for each hour close to the demand requirement.

With the new SED model (4), the task loss is updated as follows:

$$L_{t\_new}(\theta) = \sum_{t=1}^{T} \lambda_s \max\left(y_{train,t} - \sum_{g=1}^{N_g} P_{g,t}^*, 0\right)$$
$$+ \lambda_e \max\left(\sum_{g=1}^{N_g} P_{g,t}^* - y_{train,t}, 0\right) + \lambda_l \max \sum_m \Gamma_{l,m}(k_r \times y_{train,t} - \sum_{g \in \mathcal{G}(m)} P_{g,t}^*) - F_l, 0)$$
$$+ \lambda_l \max(\sum_m \Gamma_{l,m}(\sum_{g \in \mathcal{G}(m)} P_{g,t}^* - k_r \times y_{train,t}) - F_l, 0)$$
$$+ 1/2 \times \left(\sum_{g=1}^{N_g} P_{g,t}^* - y_{train,t}\right)^2 \quad (5)$$

As shown in Eq. (5), the indicator is removed so it is easy to find the derivative of task loss function (5) with respect to parameter $\theta$. After obtaining task loss $L_{t\_new}(\theta)$, there is no need to check the probabilistic constraints. Instead, $\theta$ is directly updated by taking gradient decent of task loss function (5), which largely improves calculation speed.

*D. Solution Method to the SED*

The SED model (4) is difficult to solve without knowing the distribution of $y$. In this paper, it is assumed that the forecasted load $y_i$ follows the Gaussian distribution with mean $\mu_i$ and variance $\sigma_i^2$. It should be noted that the true distribution of the forecasted load $y_i$ in practice is not Gaussian distribution and it is not easy to find the true distribution. However, the assumption of the Gaussian distribution can make it easy to solve the model and get relatively accurate results, thus it is widely used in practice. Furthermore, even if the distribution of the forecasted load $y_i$ does not strictly follow the Gaussian distribution, every time the proposed LfEdNet is trained, the $\theta$ is updated and the parameters of the distribution of predictive load $y_i$ are modified to reduce the distance from the true value so the distribution of forecasted load $y_i$ gets closer to its true distribution. Therefore, under this assumption, the simplified form of Eq. (4) is obtained and shown in (6):

$$\min_{P \in R^{T \times N_g}} f = \sum_{t=1}^{T} \alpha_t + \beta_t + C_{P,t} \quad (6)$$

$$\text{s.t. } (2.1) - (2.3)$$

where,

$$\alpha_t = \mathrm{E}_{y \sim p(y|x;\theta)}(C_{1,t}) = (\lambda_s + \lambda_e)(\sigma_t^2 PDF(s_t; \mu_t, \sigma_t^2) + (s_t - \mu_t)CDF(s_t; \mu_t, \sigma_t^2)) - \lambda_s(s_t - \mu_t) \quad (6.1)$$

$$\beta_t = \mathrm{E}_{y \sim p(y|x;\theta)}(C_{2,t}) = \lambda_l \sum_{l=1}^{N_l} (\sigma_t^2 PDF(s_{l,t}^+; \mu_t, \sigma_t^2) + (s_{l,t}^+ - \mu_t)CDF(s_{l,t}^+; \mu_t, \sigma_t^2)) + \sum_{l=1}^{N_l} [\lambda_l(\sigma_t^2 PDF(s_{l,t}^-; \mu_t, \sigma_t^2) + (s_{l,t}^- - \mu_t)CDF(s_{l,t}^-; \mu_t, \sigma_t^2)) - (s_{l,t}^- - \mu_t)] \quad (6.2)$$

$$C_{r,t} = \mathrm{E}_{y \sim p(y|x;\theta)}(C_{2,t} + C_{p,t}) = 1/2 \times \left(\left(\sum_{g=1}^{N_g} P_{g,t} - \mu_t\right)^2 + \sigma_t^2\right) + C_{p,t} \quad (6.3)$$

and

$$s_t = \sum_{g=1}^{N_g} P_{g,t} \quad (6.4)$$

$$s_{l,t}^+ = \sum_m \Gamma_{l,m}(\sum_{g \in \mathcal{G}(m)} P_{g,t}^* - F_l)/\sum_m \Gamma_{l,m}k_r \quad (6.5)$$

$$s_{l,t}^- = \sum_m \Gamma_{l,m}(\sum_{g \in \mathcal{G}(m)} P_{g,t}^* + F_l)/\sum_m \Gamma_{l,m}k_r \quad (6.6)$$

where $PDF(;\mu_t, \sigma_t^2)$ and $CDF(;\mu_t, \sigma_t^2)$ represent the Gaussian probability density function (PDF) and cumulative distribution function (CDF), respectively.

It is known that the derivative of Gaussian CDF is Gaussian PDF so the second derivative of Eq. (6) exists. The sequential quadratic programming (SQP) is a good choice to solve a problem with the second derivative. SQP is an iterative method and solves a sequence of optimization subproblems, each of which optimizes a quadratic model of the objective function [35].

If SQP is applied to the SED model (6), the compact form of the subproblem in the SQP framework is given as:

$$P^{(k+1)} = \underset{P}{\operatorname{argmin}} \frac{1}{2} P^T H P + J^T P \quad (7)$$

$$\text{s.t. } GP \leq h \quad (7.1)$$

where, $H \in \mathbb{R}^{TN_g \times TN_g}$ is Hessian matrix and $J \in \mathbb{R}^{(T \cdot N_g)}$ is Jacobian matrix. Eq. (7.1) is the compact form of Eqs. (2.1) - (2.3). The specific forms of $H$ and $J$ are described in the Appendix. The pseudo code of applying SQP to solve the SED model is described in Algorithm 1, where subproblem (7) is solved repeatedly until the convergence criterion $|P^{(k+1)} - P^{(k)}| < \varepsilon$ is met, and the optimal value $P^*(x, \theta)$ is equal to $P^{(k+1)}$ at the convergence.

**Algorithm 1** Sequential Quadratic Programming
1: Set $k = 0$, and define convergence tolerance $\varepsilon$
2: **do**
3:   Calculate $H$ and $J$
4:   Formulate subproblem (7)
5:   Solve subproblem (7) and obtain optimal $P^{(k+1)}$
6:   **if** $\left|P^{(k+1)} - P^{(k)}\right| > \varepsilon$ **then**
7:     $P^{(k)} = P^{(k+1)}$
8:     $k = k + 1$
9:   **else**
10:    **end**
11: **continue**
12: Set $P^* = P^{(k+1)}$

*E. Overall Training Process*

The complete training process of LfEdNet is shown in Fig. 4(a), where $i$ and $N_{train}$ are the index and the total number of training epochs/times, respectively. The first step is to obtain the forecasted load in Lf layer and then the optimal economic dispatch in Ed layer. The next step is to calculate the task loss by Eq. (5). The objective of the LfEdNet is to minimize task loss by updating parameters $\theta$ of the Lf layer. The task-based load forecasting model LfEdNet proposed in this paper will be compared in the case study with LfNet, a benchmark four-layer ANN model as shown in Fig. 2 before. The complete training process of the LfNet is shown in Fig. 4(b). It can be seen that LfNet is a reduced version of LfEdNet that only has the Lf layer and does not have the Ed layer. In addition, the loss functions of the two models are different. The loss function of the LfNet is Eq. (1) while LfEdNet uses Eq. (5) as its loss function.

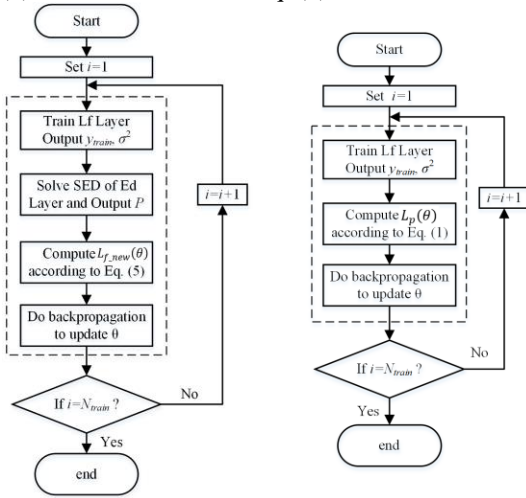

(a) Training process of LfEdNet   (b) Training process of LfNet
Fig. 4 Training process of LfEdNet and LfNet

## III. CASE STUDY

In this section, a case study is conducted to demonstrate the effectiveness of LfEdNet. First, the dataset used for the case study is explained. Then, the detailed training process of LfEdNet is presented. The results of LfEdNet are compared with those of the benchmark four-layer ANN model (i.e., LfNet) on two evaluation criteria including the accuracy of load forecasting and the SED cost of supplying the forecasted loads.

The case study is conducted for a 3-bus power system on a PC with Intel i7-9700K CPU @ 3.6 GHz, 8GB RAM, and NVIDIA GeForce RTX 2080 Ti. The settings of parameters for the LfEdNet in this case study are shown in Table I.

TABLE I SETTINGS OF PARAMETERS

| Parameters | Values |
|---|---|
| $N_{train}$ | 900 |
| $\lambda_s$ | 50 |
| $\lambda_e$ | 0.5 |
| $\lambda_l$ | 50 |
| $\varepsilon$ | $10^{-6}$ |

*A. Data Description*

The dataset used in this case study is the same as the dataset used in Ref. [33] which is a modified version of the PJM load from 2012 to 2016. When training the benchmark ANN model (LfNet), 80% of the samples is used as training dataset and the remaining 20% as the testing dataset. When training the LfEdNet model, the training dataset used in the benchmark ANN model (LfNet) is shuffled and 80% of them is used as the new training data and the remaining 20% as the validating data for hyperparameters tuning to avoid overfitting.

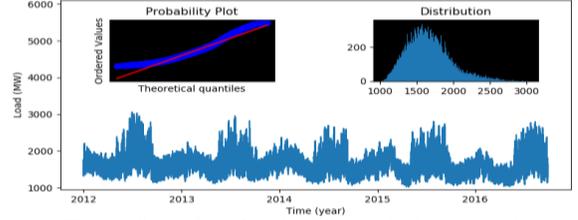

Fig. 5. Illustration of the load dataset in the case study.

Fig. 5 shows the load dataset used in the paper. The bottom part is the total load data from 2012 to 2016. Two features of the data are analyzed. The first one is the distribution of the data shown on the upper right. It shows that the distribution of the load data is not strictly Gaussian distribution but is close to Gaussian distribution. The second one is the normal probability plot shown on the upper left. The normal probability plot is a graphical technique for assessing whether or not a dataset is approximately Gaussian distribution [37]. If the points form a straight line shown as the red line in the upper left of Fig. 5, the data follow the Gaussian distribution. Otherwise, if the points deviate from the red straight line, it indicates that the data deviate from the Gaussian distribution. The blue points are the data used in this case study. The probability plot in Fig. 5 shows that the dataset nearly follows Gaussian distribution but does not exactly follow the Gaussian distribution. Therefore, the assumption in Section II.C that the predicted load $y_i$ follows the Gaussian distribution is reasonable. Before all data are fit into the LfEdNet or LfNet, they are normalized using Eq. (15).

$$x_{norm} = \frac{x - mean}{standard\_deviation} \quad (15)$$

*B. Training Process of LfEdNet*

The Lf layer of the LfEdNet is pre-trained and initialized using the complete training process of the LfNet as shown in Fig. 4(b). Fig. 6 shows the change of prediction loss during the pre-training process. After training 1000 times, the prediction loss is roughly stable at 0.005. Then the entire LfEdNet is trained using the complete training process as shown in Fig.

4(a). Fig. 7 shows the change of task loss with respect to training times. With the increase of training times, the task loss decreases. After $N_{train}$ times of training, the task loss is basically stable at around 1.9. Fig. 8 shows the change of prediction loss of the Lf layer during training LfEdNet. After $N_{train}$ times of training, the prediction loss of the LfEdNet is basically stable at around 0.0073.

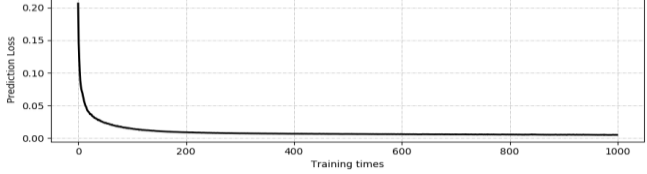

Fig. 6. The change of prediction loss of the Lf layer during the pre-training of the LfEdNet

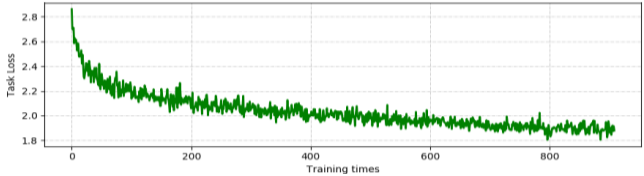

Fig. 7. The change of task loss of the LfEdNet during the training of the LfEdNet

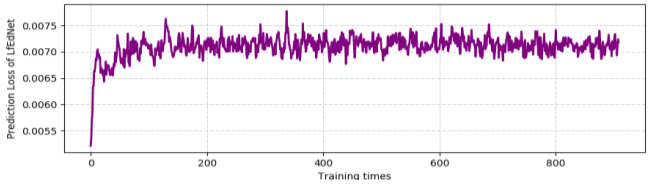

Fig. 8. The change of prediction loss of the Lf layer during the training of the LfEdNet

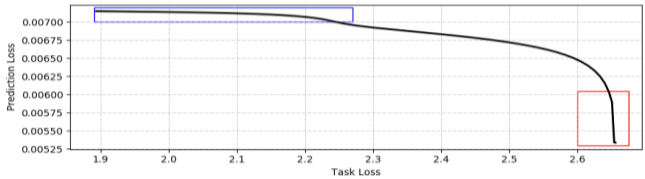

Fig. 9. The relationship between task loss and prediction loss.

Fig. 9 shows the relationship between task loss (SED cost) in the X axis and prediction loss (forecasting accuracy) in the Y axis. The prediction loss increases (lower forecasting accuracy) as the task loss decreases (lower SED cost) during the training of LfEdNet. Thus, there is a tradeoff between prediction loss and task loss for LfEdNet. If we want to decrease task loss, prediction loss increases, while if we want to reduce prediction loss, task loss cannot be the minimum one. Therefore, the prediction loss of LfEdNet is larger than that of a model without the Ed layer (i.e., LfNet). However, there is no significant gap between the prediction losses of the two models (here the difference is only 0.0023).

On the other hand, it can be observed from the blue rectangle area in Fig. 9 that for a similar prediction loss (within a difference less than 0.00025), the LfEdNet can always find the result that has minimum SED cost. Also, as is shown in the red rectangle area in Fig. 9, the task loss remains around 2.65 when the prediction loss changes from 0.006 to 0.00525. It indicates that without using the LfEdNet, the prediction loss can decrease slightly but task loss remains to be large.

Fig. 10 shows that there is a small gap between the task losses of LfEdNet on both training dataset and test dataset, which means the LfEdNet is well trained yet not over trained.

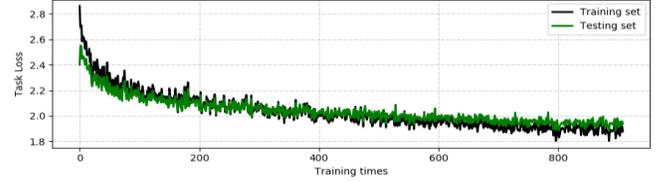

Fig. 10. Task losses of the LfEdNet on both training and test datasets

### C. LfEdNet Model vs. LfNet Model

The performance of the two models are compared from two aspects: the first one uses the mean absolute percentage error (MAPE) as evaluation criterion, which is the conventional metric for measuring the accuracy of a load forecasting model, and the second one uses the real SED cost in percentage as evaluation criterion, which represents the ultimate objective of generation scheduling. It should be noted that both tests in this part are conducted on the same test dataset to check the performance of LfEdNet. In addition, the mean and standard deviation of real SED costs for each hour of LfEdNet and LfNet are shown to compare the two models. We also calculate and compare the hourly mean values and standard deviations of the task loss of the two models. At the end, the forecasted loads for a continuous 30-day period is shown.

Table II shows the performance of the two models on MAPE and real SED cost in percentage for 10 repeated tests. It is observed that MAPE of LfEdNet is slightly larger than MAPE of LfNet for each test. The average MAPE of LfEdNet is 3.90% and that of LfNet is 3.54%. This means the forecasted load of LfNet is closer to the real load than the forecasted load of LfEdNet. When real SED cost is the evaluation criterion, LfEdNet shows its advantages. Table II column "SED Cost" shows the real SED cost in percentage with the average SED cost of the LfEdNet ($2,557,107.95) as the base value. The average SED cost of the LfNet is 100.24%, which is 0.24% higher than that of the LfEdNet. Therefore, the proposed LfEdNet model has lowered the SED cost by 0.24% at the expense of slight reduction of 0.36% in prediction accuracy.

TABLE II PERFORMANCE OF LFEDNET AND LFNET

| Test # | MAPE (%) | | SED Cost (%) | |
|---|---|---|---|---|
| | LfEdNet | LfNet | LfEdNet | LfNet |
| 1 | 4.00 | 3.43 | 100.00 | 100.26 |
| 2 | 3.88 | 3.60 | 100.00 | 100.25 |
| 3 | 3.89 | 3.70 | 100.01 | 100.26 |
| 4 | 3.87 | 3.50 | 99.99 | 100.21 |
| 5 | 3.95 | 3.51 | 100.04 | 100.26 |
| 6 | 3.90 | 3.50 | 100.00 | 100.25 |
| 7 | 4.01 | 3.64 | 100.01 | 100.25 |
| 8 | 3.83 | 3.51 | 100.00 | 100.22 |
| 9 | 3.82 | 3.51 | 99.98 | 100.27 |
| 10 | 3.86 | 3.50 | 99.97 | 100.21 |
| **Average** | 3.90 | 3.54 | 100.00 | 100.24 |

The hourly mean SED costs of the two models are shown in Fig. 11. The black and red lines show the results of LfNet and LfEdNet, respectively. It can be observed that the mean SED cost of LfEdNet is smaller than that of LfNet at each hour. It should be noted that the standard deviations of the hourly SED costs are small, so they are not visible if shown in Fig. 11.

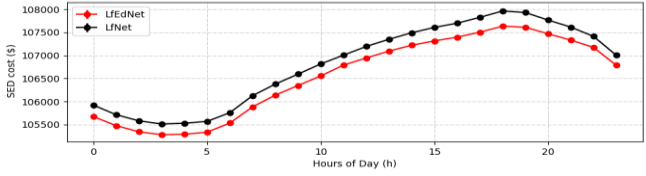

Fig. 11. Hourly SED costs of LfEdNet and LfNet

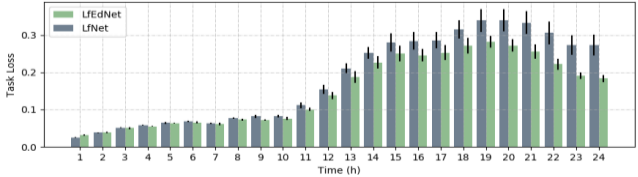

Fig. 12. Hourly task losses and standard deviations of LfEdNet and LfNet

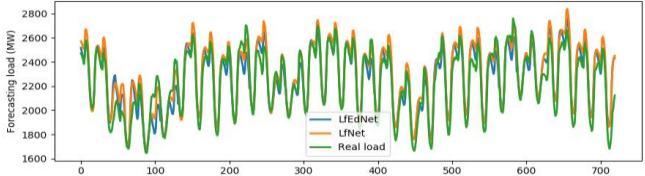

Fig. 13. Forecasted loads from LfEdNet and LfNet vs. real loads for 30 days

Results on task loss that corresponds to the second evaluation criteria are shown in Fig. 12. The gray bar shows the mean values of LfNet and the green bar shows the mean values of LfEdNet. The black line on each bar represents the standard deviation. It can be observed that the mean values of task loss of LfEdNet are smaller than those of LfNet at every hour, and the standard deviations of task loss of LfEdNet are also smaller at every hour, which means the results of LfEdNet are better and more stable. In this case, it shows that for the same dataset and the same forecasting algorithm, the forecasted load from LfEdNet can result in lower task loss than that from the LfNet. Furthermore, LfEdNet is flexible because the Lf layer can use any forecasting algorithm like LSTM, CNN, deep residual network and so forth. No matter what forecasting algorithm is used, the forecasting results of LfEdNet always result in a lower cost of SED than that of a model-free load forecasting model using the same forecasting algorithm.

Fig. 13 shows the forecasted load of the two models for 30 days. The real load, the forecasted load of LfEdNet, and the forecasted load of LfNet are shown in green line, blue line, and orange line, respectively. It can be observed that the forecasted load of these two models are close to the test data. LfNet performs better than LfEdNet because its objective is to minimize prediction loss. However, from Table II and also Fig. 13, it can be observed that the average MAPE of the two models does not have a big gap (the difference is only 0.5%) and both forecasted loads are near the real values. Therefore, the forecasting results of LfEdNet are relatively good and acceptable.

## IV. CONCLUSION

This paper proposes LfEdNet, a task-based day-ahead load forecasting model. LfEdNet combines load forecasting and day-ahead SED together in one model. A four-layer ANN is used for load forecasting in the Lf layer. The day-ahead SED model of the Ed layer is to minimize the SED cost and obtain the optimal economic dispatch. The objective of LfEdNet is to minimize the loss function so that the forecasted load can always find lower SED cost than other model-free forecasting models. Therefore, the proposed LfEdNet can output a relatively accurate load forecast that minimizes the SED cost.

The case studies first analyze the distribution of the dataset and shows the data used in this paper nearly follows Gaussian distribution. The case studies demonstrate that the proposed LfEdNet has lower SED cost than the model-free LfNet and also has relatively good load forecasting accuracy, although at the expense of slight reduction in prediction accuracy.

Although LfEdNet has advantages, there are still limitations for LfEdNet. The time complexity of Ed layer is cubic time complexity with respect to the number of variables. Due to this, the dimension of input variables is limited to less than 1000. In this case, the model proposed in this paper is suitable for a power system that is less than 41 buses. Therefore, LfEdNet at the current stage is only suitable for small-scale power systems, such as a microgrid. One of our future work will be focused on how to extend it for large-scale power systems. Furthermore, the model only considers economic dispatch problem but does not consider the unit commitment problem, so another future work will be on how to apply the proposed model to a unit commitment problem.

## APPENDIX

In this part, the details for Eq. (8) is presented.

$$P^{(k+1)} = \underset{P}{\operatorname{argmin}} \frac{1}{2} P^T H P + J^T P \quad (8)$$

$$\text{s.t. } GP \leq h \quad (8.1)$$

where, the expression of $f$ is shown in Eq. (6)

$$\min_{P \in R^{T \times N_g}} f = \sum_{t=1}^{T} \alpha_t + \beta_t + C_{P,t} \quad (6)$$

$H \in \mathbb{R}^{TN_g \times TN_g}$ is Hessian matrix shown in Eq. (A-1),

$$H = \begin{pmatrix} \frac{\partial^2 f}{\partial P_{1,1}^2} & \cdots & \frac{\partial^2 f}{\partial P_{1,1} P_{N_g,T}} \\ \vdots & \ddots & \vdots \\ \frac{\partial^2 f}{\partial P_{N_g,T} P_{1,1}} & \cdots & \frac{\partial^2 f}{\partial P_{N_g,T}^2} \end{pmatrix} \quad (A-1)$$

$J \in \mathbb{R}^{(T \cdot N_g)}$ is Jacobian matrix expressed in Eq. (A-2)

$$J = \nabla f - P^T H \quad (A-2)$$

The first derivative and the second derivative of each part of function $f$ are expression on Eqs. (A-3)-(A-12) in details.

$$\nabla f = \begin{pmatrix} \frac{\partial f}{\partial P_{1,1}} & \cdots & \frac{\partial f}{\partial P_{N_g,T}} \end{pmatrix} \quad (A-3)$$

$$\nabla f = \sum_{t=1}^{T} \nabla \alpha_t + \nabla \beta_t + \nabla C_{P,t} \quad (A-4)$$

$$\frac{\partial f}{\partial P_{g,t}} = \sum_{t=1}^{T} \frac{\partial \alpha_t}{\partial s_t} \frac{\partial s_t}{\partial P_{g,t}} + \frac{\partial \beta_t}{\partial s_{l,t}^+} \frac{\partial s_{l,t}^+}{\partial P_{g,t}} + \frac{\partial \beta_t}{\partial s_{l,t}^-} \frac{\partial s_{l,t}^-}{\partial P_{g,t}} + \frac{\partial C_{P,t}}{\partial P_{g,t}} \quad (A-5)$$

$$\frac{\partial^2 f}{\partial P_{g,j} P_{m,n}} = \sum_{j=1}^{T} \frac{\partial^2 \alpha_j}{\partial s_j^2} \frac{\partial s_t}{\partial P_{m,n}} \frac{\partial s_t}{\partial P_{g,t}} + \frac{\partial^2 \beta_t}{\partial s_{l,t}^{+\,2}} \frac{\partial s_{l,t}^+}{\partial P_{m,n}} \frac{\partial s_{l,t}^+}{\partial P_{g,t}} +$$
$$\frac{\partial^2 \beta_t}{\partial s_{l,t}^{-\,2}} \frac{\partial s_{l,t}^-}{\partial P_{m,n}} \frac{\partial s_{l,t}^-}{\partial P_{g,t}} + \frac{\partial^2 C_{P,t}}{\partial P_{g,t} P_{m,n}} \quad (A-6)$$

$$\frac{\partial \alpha_t}{\partial s_t} = (\lambda_s + \lambda_e) CDF(s_t; \mu_t, \sigma_t^2) - \lambda_s \quad (A-7)$$

$$\frac{\partial^2 \alpha_t}{\partial s_t^2} = (\lambda_s + \lambda_e) PDF(s_t; \mu_t, \sigma_t^2) \quad (A-8)$$

$$\frac{\partial \beta_t}{\partial s_{l,t}^+} = \sum_{l=1}^{N_l} \lambda_l CDF\left(s_{l,t}^+; \mu_t, \sigma_t^2\right) \quad \text{(A-9)}$$

$$\frac{\partial \beta_t}{\partial s_{l,t}^-} = \sum_{l=1}^{N_l} \lambda_l CDF\left(s_{l,t}^-; \mu_t, \sigma_t^2\right) - \lambda_l \quad \text{(A-10)}$$

$$\frac{\partial^2 \beta_t}{\partial s_{l,t}^{+\,2}} = \sum_{l=1}^{N_l} \lambda_l PDF\left(s_{l,t}^+; \mu_j, \sigma_t^2\right) \quad \text{(A-11)}$$

$$\frac{\partial^2 \beta_j}{\partial s_{l,t}^{-\,2}} = \sum_{l=1}^{N_l} \lambda_l PDF\left(s_{l,t}^-; \mu_t, \sigma_t^2\right) \quad \text{(A-12)}$$

REFERENCES

[1] S. Fallah, M. Ganjkhani, S. Shamshirband, and K. Chau, "Computational intelligence on short-term load forecasting: A methodological overview," *Energies,* vol. 12, no. 3, pp.393, Jan. 2019.

[2] A. Singh, Ibraheem, S. Khatoon, M. Muazzam, and D. Chaturvedi, "Load forecasting techniques and methodologies: A review," in *2012 2nd International Conference on Power, Control and Embedded Systems,* Allahabad, pp. 1-10, Dec. 2012.

[3] G. Gross and F. Galiana, "Short-term load forecasting," *Proceedings of the IEEE,* vol. 75, no. 12, pp. 1558-1573, Dec. 1987.

[4] M. Ibrahim and S. Rahman, "Analysis and evaluation of five short-term load forecasting techniques," *IEEE Transactions on power systems,* vol. 4, no.4, pp. 1484-1491, Nov. 1989.

[5] J. Che and J. Wang, "Short-term load forecasting using a kernel-based support vector regression combination model," *Applied energy,* vol. 132, pp. 602-609, Nov. 2014.

[6] G. Mbamalu and M. El-Hawary, "Load forecasting via suboptimal seasonal autoregressive models and iteratively reweighted least squares estimation," *IEEE Transactions on Power Systems*, vol. 8, no. 1, pp. 343-348, Feb. 1993.

[7] P. Dash, H. Satpathy, A. Liew, and S. Rahman, "A real-time short-term load forecasting system using functional link network," *IEEE Transactions on Power Systems*, vol. 12, no. 2, pp. 675-680, May 1997.

[8] G. Juberias, R. Yunta, J. Moreno, and C. Mendivil, "A new ARIMA model for hourly load forecasting," in *1999 IEEE Transmission and Distribution Conference* (Cat. No. 99CH36333), New Orleans, LA, USA, vol.1, pp. 314-319, 1999.

[9] H. Nie, G., Liu, X. Liu, and Y. Wang, "Hybrid of ARIMA and SVMs for short-term load forecasting," *Energy Procedia*, vol.16, pp.1455-1460, Jan. 2012.

[10] A. Petrică, S. Stancu, and A. Tindeche, "Limitation of ARIMA models in financial and monetary economics," *Theoretical & Applied Economics,* vol. 23, no. 4, Dec. 2016.

[11] S. Fan and L. Chen, "Short-term load forecasting based on an adaptive hybrid method," *IEEE Transactions on Power Systems*, vol. 21, no. 1, pp. 392-401, Feb. 2006.

[12] H. Jiang, Y. Zhang, E. Muljadi, J. Zhang, and D. Gao, "A short term and high-resolution distribution system load forecasting approach using support vector regression with hybrid parameters optimization," *IEEE Transactions on Smart Grid*, vol. 9, pp. 3341- 3350, Nov. 2018.

[13] Z. Ma, C. Zhang, and C. Qian, "Review of machine learning in power system," in *2019 IEEE Innovative Smart Grid Technologies - Asia (ISGT Asia)*, Chengdu, China, pp. 3401-3406, 2019.

[14] Z. Deng, B. Wang, Y. Xu, T. Xu, C. Liu, and Z. Zhu, "Multi-scale convolutional neural network with time-cognition for multi-step short-term load forecasting," *IEEE Access*, vol. 7, pp. 88058-88071, Jul. 2019.

[15] H. Mori, N. Kosemura, K. Ishiguro, and T. Kondo, "Short-term load forecasting with fuzzy regression tree in power systems," in *2001 IEEE International Conference on Systems, Man and Cybernetics*, Tucson, AZ, USA, vol.3, pp. 1948-1953, 2001.

[16] C. Guan, P. Luh, L. Michel, and Z. Chi, "Hybrid Kalman filters for very short-term load forecasting and prediction interval estimation," *IEEE Transactions on Power Systems*, vol. 28, no. 4, pp. 3806-3817, Nov. 2013.

[17] L. Yin, Z. Sun, F. Gao, and H. Liu, "Deep forest regression for short-term load forecasting of power systems," *IEEE Access*, vol. 8, pp. 49090-49099, Mar. 2020.

[18] Y. Bengio, A. Courville, and P. Vincent, "Representation learning: A review and new perspectives," *IEEE Transactions on Pattern Analysis and Machine Intelligence*, vol. 35, no. 8, pp. 1798-1828, Aug. 2013.

[19] K. He, X. Zhang, S. Ren, and J. Sun, "Deep residual learning for image recognition," In *Proceedings of the IEEE Conference on Computer Vision and Pattern Recognition (CVPR)*, Las Vegas, NV, pp. 770-778, 2016.

[20] D. Amodei, S. Ananthanarayanan, R. Anubhai, J. Bai, E. Battenberg, C. Case, J. Casper, B. Catanzaro, Q. Cheng, G. Chen, and J. Chen, "Deep speech 2: End-to-end speech recognition in English and Mandarin," In *International conference on machine learning,* pp. 173-182, Jun. 2016.

[21] A. Graves and N. Jaitly, "Towards end-to-end speech recognition with recurrent neural networks," In *International conference on machine learning*, vol. 14, pp. 1764–1772, Jan. 2014.

[22] T. Wang, D. Wu, A. Coates, and A. Ng, "End-to-end text recognition with convolutional neural networks," In *Proceedings of the 21st International Conference on Pattern Recognition (ICPR2012),* Tsukuba, pp. 3304–3308, 2012.

[23] W. Kong, Z. Dong, Y. Jia, D. Hill, Y. Xu, and Y. Zhang, "Short-term residential load forecasting based on LSTM recurrent neural network," *IEEE Transactions on Smart Grid*, vol. 10, no. 1, pp. 841-851, Jan. 2019.

[24] K. Chen, K. Chen, Q. Wang, Z. He, J. Hu, and J. He, "Short-term load forecasting with deep residual networks," *IEEE Transactions on Smart Grid*, vol. 10, no. 4, pp. 3943-3952, Jul. 2019.

[25] C. Feng, M. Sun, and J. Zhang, "Reinforced deterministic and probabilistic load forecasting via Q-Learning dynamic model selection," *IEEE Transactions on Smart Grid*, vol. 11, no. 2, pp. 1377-1386, Mar. 2020.

[26] Y. Yang, W. Li, T. Gulliver, and S. Li, "Bayesian deep learning based probabilistic load forecasting in smart grids," *IEEE Transactions on Industrial Informatics,* vol. 16, no. 7, pp. 4703-4713, Jul. 2020.

[27] X. Dong, L. Qian, and L. Huang, "A CNN based bagging learning approach to short-term load forecasting in smart grid," in *2017 IEEE SmartWorld, Ubiquitous Intelligence & Computing, Advanced & Trusted Computed, Scalable Computing & Communications, Cloud & Big Data Computing, Internet of People and Smart City Innovation (SmartWorld/SCALCOM/UIC/ATC/CBDCom/IOP/SCI)*, San Francisco, CA, pp. 1-6, 2017.

[28] J. Lu, Q. Zhang, Z. Yang, and M. Tu, "A hybrid model based on convolutional neural network and long short-term memory for short-term load forecasting," in *2019 IEEE Power & Energy Society General Meeting (PESGM)*, Atlanta, GA, USA, pp. 1-5, 2019.

[29] Z. Deng, B. Wang, Y. Xu, T. Xu, C. Liu, and Z. Zhu, "Multi-scale convolutional neural network with time-cognition for multi-step short-term load forecasting," *IEEE Access*, vol. 7, pp. 88058-88071, 2019.

[30] A. Brandon and J. Kolter, "Optnet: Differentiable optimization as a layer in neural networks," In *Proceedings of the 34th International Conference on Machine Learning,* vol. 70, pp. 136-145, Aug. 2017.

[31] Y. Wu, M. Schuster, Z. Chen, Q. Le, M. Norouzi, W. Macherey, M. Krikun, Y. Cao, Q. Gao, K. Macherey, and J. Klingner, "Google's neural machine translation system: Bridging the gap between human and machine translation," *arXiv preprint arXiv*:1609.08144 , 2016.

[32] S. Ioffe and C. Szegedy, "Batch normalization: Accelerating deep network training by reducing internal covariate shift," *arXiv preprint arXiv*:1502.03167, 2015.

[33] P. Donti, B. Amos, and J. Kolter, "Task-based end-to-end model learning in stochastic optimization," In *Advances in Neural Information Processing Systems*, pp. 5484-5494, 2017.

[34] A. Nemirovski and A. Shapiro, "Convex approximations of chance constrained programs," *SIAM Journal on Optimization*, vol. 17, no. 4, pp. 969-96, 2007.

[35] J. Bonnans, J. Gilbert, C. Lemaréchal, and C. Sagastizábal, "Local methods for problems with equality and inequality constraints," In *Numerical Optimization,* 2nd ed. New York, NY, USA: Springer, 2006, ch. 15, sec. 1, pp. 256–259.

[36] PyTorch, https://pytorch.org/.

[37] J. Chambers, W. Cleveland, B. Kleiner, and P. Tukey, Graphical Methods for Data Analysis, Wadsworth, 1983.